\documentclass[showpacs,amssymb,twocolumn,floatfix,aps,notitlepage]{revtex4-1}

\usepackage{amssymb,amsfonts,amsmath,bm}
\usepackage{graphics,graphicx}
\newcommand{\mueff}{\mu_{\text{eff}} }  
\newcommand{\zcross}{z_{\text{cross}} }

\begin{document}

\title{A correlation-hole approach to the electric double layer
with counter-ions only}

\author{Ivan Palaia}
\affiliation{LPTMS, CNRS, Univ. Paris-Sud, Universit\'{e} Paris-Saclay,
91405 Orsay, France }
\author{Martin Trulsson}
\affiliation{Theoretical Chemistry, Lund University, Lund, Sweden}
\author{Ladislav \v{S}amaj}
\affiliation{Institute of Physics, Slovak Academy of Sciences, Bratislava, Slovakia} 
\author{Emmanuel Trizac}
\affiliation{LPTMS, CNRS, Univ. Paris-Sud, Universit\'{e} Paris-Saclay,
91405 Orsay, France}

\begin{abstract}
We study a classical system of identically charged counter-ions near a planar 
wall carrying a uniform surface charge density.
The equilibrium statistical mechanics of the system depends on 
a single dimensionless coupling parameter.
A new self-consistent theory of the correlation-hole type is proposed which leads 
to a modified Poisson-Boltzmann integral equation for the density profile,
convenient for analytical progress and straightforward to solve numerically.
The exact density profiles
are recovered in the limits of weak and strong couplings. 
In contrast to previous theoretical attempts of the test-charge family,
the density profiles fulfill the contact-value theorem  at all values of the coupling constant, and exhibit
the mean-field decay at asymptotically large distances from the wall, 
as expected. We furthermore show that the density corrections
at large couplings exhibit the proper dependence on
coupling parameter and distance to the charged wall.
The numerical results for intermediate values of the coupling provide 
accurate density profiles which are in good agreement with those obtained by 
Monte-Carlo simulations. The crossover to mean-field behavior at large distance
is studied in detail.
\end{abstract}

\date{\today} 

\maketitle

\section{Introduction}
Experiments with large macromolecules are often performed in water,
which is a polar solvent. This is the case for many applications using colloids, including the proteins in our bodies.
This results in the release of low valence micro-ions 
into the solution, so that the colloids acquire a surface charge density, opposite
to the charge of mobile micro-ions (coined as ``counter-ions''). The total 
surface charge can exceed thousands of elementary charges $e$.
In the first approximation, the curved surface of a macromolecule can be
replaced by an infinite rectilinear plane.

The charged macromolecule and the surrounding counter-ions form a neutral
electric double layer, see reviews \cite{Attard96,Hansen00,Messina09}.
In turn, the double layer is paramount in mediating the effective
interactions between charged bodies in solution.
At large enough Coulombic coupling, it is for instance known that like-charged macromolecules can effectively
attract each other in some intermediate distance range, as was observed experimentally
\cite{Khan85,Kjellander88,Bloomfield91,Rau92,Kekicheff93,Dubois98} 
and by computer simulations \cite{Gulbrand84,Kjellander84,Bratko86,Gronbech97}.

In a wealth of natural or synthetic systems, micro-ions can be of both signs,
with positively and negatively 
charged species.
In this paper, we restrict ourselves to simplified so-called salt-free (or deionized) 
Coulomb systems with counter-ions only. This is a convenient starting 
point for analytical progress, 
where detailed computer simulation results
are also available 
\cite{Belloni00,HoKP00,Netz00,MoreiraNetz00,MoreiraNetz01,Levin02,Burak04,Chen,Santangelo06,WLST08,Dent08,Hatlo10,TeTr15,SPKK16,Samaj16}.
Such models apply to deionized suspensions, see e.g. the experiments reported
in Refs. \cite{RaLi00,Palberg04,BDvGB04,HQCS06,RCLS08}. 
In the deionized limit, systems of counter-ions near charged surfaces
have poor screening properties, but the standard Coulomb sum rules 
relating the one-body and two-body densities do apply 
\cite{Martin88,Samaj13}.

For the system of counter-ions near a charged wall,
the high-temperature (weak-coupling, WC) limit is described by the
Poisson-Boltzmann (PB) mean-field theory \cite{Andelman06} and
by its systematic improvement via the loop expansion
\cite{Attard88,Podgornik90,Netz00}.
The opposite strong-coupling (SC) limit was investigated within 
a field-theoretical formulation of the model by using a renormalized expansion 
of virial type \cite{Shklovskii99,MoreiraNetz00,BaDH00,Goles00,MoreiraNetz01,Naji05}.
In the leading SC order and in the present planar geometry, 
the counter-ions effectively behave as non-interacting objects, as far as one is not interested in 
the tail of the density profile; this fact was confirmed numerically in
a number of numerical studies \cite{Naji,Kanduc07,Kanduc08,Jho08,Dean09,Kanduc10}.
The first correction to the single-particle density profile, calculated 
within a fugacity expansion with a renormalization of infrared divergences
\cite{MoreiraNetz01}, is correct in its functional form, but with a wrong prefactor,
departing by orders of magnitude from its Monte-Carlo (MC) estimate \cite{Santangelo06}.
Other SC approaches \cite{Perel} emphasize the two-dimensional
Wigner crystallization of mobile charges at the wall surface for low
temperatures.
Recently \cite{Samaj11}, by a perturbative approach around 
the Wigner crystal, the single particle treatment was recovered in 
the leading SC order.
Moreover, the derived prefactor of the first SC correction is in excellent 
agreement with MC simulations, also in the coupling range where no 
Wigner crystal is formed (strongly modulated liquid regime). Noteworthy are also field theoretic techniques,
that allow to cover the crossover regime between WC and SC,
by a proper splitting of the interactions between ions, discriminating short and long
distances \cite{Chen,Santangelo06,HaLu09,Hatlo10}.

For a system of identical charges with Coulomb repulsion, the pair
correlation function is strongly depleted at small distances.
This gives credit to the image of a correlation hole around each ion in the system, 
an idea that turned useful in various approaches going beyond the PB theory
\cite{Nordholm84,Rouzina96,BaDH00,Fors04,Santangelo06,Chen,Hatlo10,Bakhshandeh11,Varenna}. 
Recently \cite{Samaj16}, for a dielectric interface, the single particle 
strong-coupling view was combined with the idea of the correlation hole,
to obtain very accurate density profiles for strongly to moderately coupled 
charged fluids. This latter contribution provides the most accurate theory available so far for these systems.
We emphasize that this approach is not self-consistent, and does not reproduce mean-field PB results at small
couplings, two key differences with the 
theory to be developed below.

In Ref. \cite{Burak04}, an attempt has been made to establish a universal 
theory which works adequately for any value of the coupling. 
Based on a mean-field treatment of the ions response to the presence 
of a test charge, the exact density profile was reproduced in the limits
of weak and strong couplings.
For intermediate values of the coupling, the obtained approximate density
profiles agree with MC simulations, except for two shortcomings.
Firstly, the contact theorem for the counter-ion number density at the wall 
\cite{Henderson,Carnie81,Wennerstrom82} is not satisfied.
Secondly, although a crossover from exponential to algebraic decay
is observed at large distances from the wall, there is an additional 
prefactor to the mean-field PB solution which depends on the coupling 
constant. 
This is in contradiction with the common expectation that mean-field
should hold at large distances from the wall
\cite{Shklovskii99,MoreiraNetz01,Chen,Santangelo06,Santos09,Varenna},
as the small density of counter-ions should effectively drive the system into the WC regime.
Note that the loop corrections to the PB solution 
\cite{Attard88,Podgornik90,Netz00} are consistent with this expectation.

In this work, we propose a self-consistent theory for counter-ions 
near a charged rectilinear wall, which is based on the idea of 
a cylindrical correlation hole.
As was the case in the test-charge approach of Ref. \cite{Burak04}, 
the exact density profiles are recovered in the limits of weak and 
strong couplings. 
But in contrast to that theory, at all values of the coupling constant do
the density profiles fulfill the contact-value theorem. 
Moreover, the density profiles are exactly of mean-field type at asymptotically
large distances from the wall, as expected. This allows us to address the elusive 
question of the asymptotic large distance crossover to mean-field
in this geometry.

The article is organized as follows.
In Sec. \ref{II}, we introduce the basic notations for the model .
The correlation-hole approach is presented in Sec. \ref{III}.
For the sake of analogy and completeness, the derivation of the PB theory is 
provided as well. Analytical progress was made possible by an original 
rederivation of the contact theorem, that does not require the explicit resolution
of the theory under study.
Section \ref{IV} derives a number of exact results. The SC limit is
worked out. Then, at arbitrary coupling parameter, the large-distance behavior of the density profile
is shown to be exactly of the PB mean-field type.
In addition, we derive the subleading contribution to the mean-field tail. 
Numerical results for the density profile at specific values of 
the coupling constant are compared with those obtained by the test-charge
method \cite{Burak04} and by MC simulations in Sec. \ref{VI}. 
The crossover distance from the wall to the mean-field algebraic decay
of the density profile is determined too.
A short recapitulation and concluding remarks are given in Sec. \ref{VII},
where we present some results pertaining to an interacting two-plate system
 both in Monte Carlo and
within our self-consistent scheme.

\section{Basic formalism} \label{II}
\begin{figure}[tbp]
\begin{center}
\includegraphics[width=0.2\textwidth,clip]{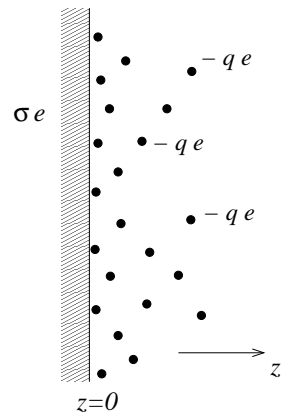}
\caption{The electric double layer with counter-ions of valence $q$. The interface
at $z=0$ bears a surface charge $\sigma e$, so that the system as a whole is electroneutral.}
\label{fig:geometry}
\end{center}
\end{figure}

We consider the one-wall geometry pictured in Fig. \ref{fig:geometry}, with positions
denoted by ${\bf r}=(x,y,z)$. 
A hard wall, impenetrable 
to particles, is localized in the half-space $\{ {\bf r},z<0\}$. 
In the complementary half-space $\{ {\bf r},z>0\}$, there are $N$ 
mobile $q$-valent counter-ions (classical point-like particles) of charge
$-q e$, where $e$ is the elementary charge. 
The particles are immersed in a solution with the same 
dielectric constant $\varepsilon$ as the confining wall, so that no
electrostatic image forces ensue. 
The infinite wall surface, localized at $z=0$, carries 
a fixed uniform surface-charge density $\sigma e$ with $\sigma>0$. 
The system as a whole is electro-neutral, 
and the particles are in thermal equilibrium at some inverse temperature
$\beta = 1/(k_{\rm B}T)$. 

There are two relevant length scales in the model.
In Gauss units, two unit charges at distance $r$ interact by the 3D Coulomb 
energy $e^2/(\varepsilon r)$; the distance at which this energy coincides with
the thermal energy $k_{\rm B}T$ is the Bjerrum length
\begin{equation}
\ell_{\rm B} = \frac{\beta e^2}{\varepsilon} .
\end{equation}
The potential energy of an isolated counter-ion of charge $-q e$ at 
distance $z$ from the wall surface is given by
\begin{equation}
E(z) = \frac{2\pi q e^2\sigma}{\varepsilon} z ;
\end{equation}
the distance at which this energy equals to the thermal energy $k_{\rm B}T$ 
defines the Gouy-Chapman length
\begin{equation}
\mu \,=\, \frac{1}{2\pi q \ell_{\rm B}\sigma} .
\label{eq:gc}
\end{equation}
The dimensionless coupling parameter $\Xi$, reflecting the strength
of electrostatic correlations, is defined as the ratio of the two
length scales:
\begin{equation}
\Xi \, = \, \frac{q^2 \ell_{\rm B}}{\mu} = 2\pi q^3 \ell_{\rm B}^2 \sigma .
\end{equation}

Denoting by $\langle \cdots \rangle$ the canonical thermal average, 
the particle number density at point ${\bf r}$ (with thus $z\geq 0$) is defined as 
$n({\bf r}) = \langle \sum_{i=1}^N \delta({\bf r}-{\bf r}_i) \rangle$.
It depends only on the distance $z$ from the wall, $n({\bf r}) = n(z)$.
The electroneutrality condition 
corresponds to the constraint
\begin{equation} \label{electroneutral2}
q \int_0^{\infty} dz \, n(z) = \sigma .
\end{equation}
The contact density of counter-ions at the wall is related to the surface
charge density via the planar contact-value theorem 
\cite{Henderson,Carnie81,Wennerstrom82} as follows
\begin{equation} \label{contact1}
n(0) \, = \, 2\pi \ell_{\rm B}\sigma^2 .
\end{equation}
The averaged particle density will be often written in a rescaled form 
with a dimensionless particle $z$-coordinate considered in units of 
the Gouy-Chapman length $\mu$: 
\begin{equation} \label{ntilde}
\widetilde{n}(z) \equiv  \frac{n(\mu z)}{2\pi\ell_{\rm B}\sigma^2} .
\end{equation}
In terms of $\widetilde{n}$, the electroneutrality requirement
(\ref{electroneutral2}) and the contact-value constraint (\ref{contact1})
take the forms
\begin{equation} \label{normalization}
\int_0^{\infty} dz \,\widetilde{n}(z) = 1
\end{equation} 
and
\begin{equation} \label{contact}
\widetilde{n}(0) = 1 ,
\end{equation}
respectively.
To avoid unnecessarily heavy notations, $z$ will in the remainder refer to the 
rescaled distance $z/\mu$, whenever it appears in a expression involving the reduced
density $\widetilde n$.

The model is exactly solvable in two limits.
In the weak-coupling limit $\Xi\to 0$, the PB approach \cite{Andelman06} 
implies a slowly decaying particle density profile
\begin{equation} \label{PBsolution}
\widetilde{n}_{\rm PB}(z) = \frac{1}{(1+z)^2} .
\end{equation}
In the strong-coupling limit $\Xi\to\infty$, the single-particle picture
of counter-ions in the linear surface charge potential 
\cite{MoreiraNetz00,MoreiraNetz01,Varenna} leads to an exponentially decaying profile
\begin{equation} \label{SCsolution}
\widetilde{n}_{\rm SC}(z) = \exp(-z) .
\end{equation}

\section{The correlation-hole approach} \label{III}
At any point ${\bf r}$ with $z\geq 0$, the relation between the (averaged) 
electric potential $\psi$ and the charge distribution $\rho$ is given by 
the Poisson equation
\begin{equation}
\nabla^2 \psi({\bf r}) = - \frac{4\pi}{\varepsilon} \rho({\bf r}) .
\end{equation}
For the present geometry, the electrostatic potential and the charge 
distribution $\rho = -q e n$ depend only on the distance from the wall
$z$, so that
\begin{equation}
\frac{d^2}{dz^2}\psi(z) = \frac{4\pi}{\varepsilon} q e n(z) .
\end{equation}
With respect to the boundary condition for the electric field at $z=0$,
\begin{equation}
\frac{d}{dz} \psi(z) = - \frac{4\pi}{\varepsilon} \sigma e 
\end{equation}
and the 1D relation
\begin{equation}
\frac{d^2}{dz^2} \frac{\vert z\vert}{2} = \delta(z)
\end{equation}
with $\delta$ the Dirac delta distribution, the electric potential 
is expressible explicitly as
\begin{equation} \label{elpot}
\psi(z) \,=\,  - \frac{2\pi}{\varepsilon} \sigma e z \,+\, 
\frac{2\pi}{\varepsilon} q e \int_0^{\infty} dz' \, (\vert z-z'\vert - z') \,n(z') .  
\end{equation}
The interpretation of this expression is transparent: in addition to 
the bare plate potential (first term on the rhs, linear in $z$),
the mobile counter-ions contribute to the electric potential though the
integral term.
The potential is determined up to an irrelevant constant; here we fixed 
the ``gauge'' $\psi(0) = 0$.

\subsection{PB theory}
The electrostatic energy of a counter-ion in the potential $\psi(z)$ is
$-q e \psi(z)$.
In the PB approach, the particle density is related locally to 
the corresponding Boltzmann factor as 
\begin{equation} \label{PBrelation}
n_{\rm PB}(z) = n_0 \exp\left[ \beta q e \psi_{\rm PB}(z) \right]
\end{equation}
where the parameter $n_0$ ensures the normalization (\ref{electroneutral2}).
In terms of the dimensionless $\widetilde{n}$ (\ref{ntilde}), 
the self-consistent PB equation (\ref{PBrelation}) is written as
\begin{equation} \label{PBfinal}
\widetilde{n}_{\rm PB}(z) = \widetilde{n}_0 \exp\left[ \phi_{\rm PB}(z) \right] ,  
\end{equation}
where the PB reduced potential $\phi_{\rm PB}$ is given by
\begin{equation} \label{phiPB}
\phi_{\rm PB}(z) = -z + \int_0^{\infty} dz' 
(\vert z-z'\vert - z') \widetilde{n}_{\rm PB}(z') .  
\end{equation}    
Note the gauge $\phi_{\rm PB}(0) = 0$. 

The normalization constant $\widetilde{n}_0$ is determined by the
electroneutrality condition (\ref{normalization}) through
\begin{equation} \label{n0}
\widetilde{n}_0 = \frac{1}{\int_0^{\infty} {\rm d}z 
\exp\left[ \phi_{\rm PB}(z) \right]} .
\end{equation}
There exists a simple way to obtain the explicit value of $\widetilde{n}_0$;
it will prove useful below and we thus present it in its simplest clothing.
We first differentiate the $\phi_{\rm PB}$-potential (\ref{phiPB}) 
with respect to $z$:
\begin{equation} \label{derphi}
\frac{d}{dz} \phi_{\rm PB}(z) =  - 1 + \int_0^{\infty} dz' \widetilde{n}_{\rm PB}(z')  
{\rm sgn}(z-z') ,  
\end{equation}  
where sgn denotes the standard signum (sign) function.
The integral
\begin{eqnarray}
\int_0^{\infty} dz \left( \frac{d \phi_{\rm PB}}{dz} +1 \right) 
\widetilde{n}_{\rm PB}(z) = \phantom{aaaaaaaaaaaa} \nonumber \\  
\int_0^{\infty} dz \int_0^{\infty} dz' \widetilde{n}_{\rm PB}(z) 
\widetilde{n}_{\rm PB}(z') {\rm sgn}(z-z') \label{nul} 
\end{eqnarray}
vanishes due to the anti-symmetric property of the function under integration in the rhs
with respect to the interchange transformation $z\leftrightarrow z'$.
From (\ref{PBfinal}) we get
\begin{equation}
\widetilde{n}_{\rm PB}(z) \frac{d \phi_{\rm PB}}{dz} 
= \frac{d \widetilde{n}_{\rm PB}(z)}{d z} . 
\end{equation}
Consequently, we have from (\ref{nul}) that
\begin{equation}
\int_0^{\infty} dz \frac{d}{dz} \widetilde{n}_{\rm PB}(z)
= - \int_0^{\infty} dz \widetilde{n}_{\rm PB}(z) .  
\end{equation}
The density $\widetilde{n}(z)$ vanishes as $z\to\infty$, so that
\begin{equation} \label{PBcontact}
\widetilde{n}_{\rm PB}(0) = \int_0^{\infty} dz \widetilde{n}_{\rm PB}(z) = 1 ,
\end{equation}
which is nothing but the contact-value theorem (\ref{contact}). 
We see that, within the PB theory, the normalization (\ref{normalization})
automatically ensures the contact-value theorem (\ref{contact}), 
and vice versa.
Under the gauge $\phi_{\rm PB}(0) = 0$, the contact-value relation
(\ref{PBcontact}) fixes 
\begin{equation}
\widetilde{n}_0 = 1
\end{equation}
in (\ref{PBfinal}).
 It is easy to check that under this normalization, the PB solution 
(\ref{PBsolution}) satisfies Eqs. (\ref{PBfinal}) and (\ref{phiPB}).

\subsection{Inclusion of the correlation hole}
In the single-particle SC solution (\ref{SCsolution}), the only acting
potential is due to the fixed surface-charge density; this potential 
is present also in the PB solution (\ref{PBfinal}), but there is 
an additional potential related to the mean particle density profile.
Thus in some sense, the SC solution is simpler than its mean-field 
counterpart, since mutual counter-ions interactions do not contribute
to the leading order SC response. Yet, for large $\Xi$, counter-ions are strongly correlated 
in the $(x,y)$ plane, because of their strong mutual repulsion;
this leads to a marked correlation hole (``Coulomb hole''),
inaccessible to other charged particles \cite{MoreiraNetz01}.  
At smaller $\Xi$ value, the correlation hole is less marked
(in the sense that the pair correlation function does not vanish 
for distances smaller than the hole size \cite{MoreiraNetz01}),
but a feature of depletion remains. 
In addition, the form of the correlation hole depends on the distance from the wall of the particle 
under consideration.
It is expected, in the large $\Xi$ regime, that the correlation hole is cylindrical 
if the particle is close to the wall, and spherical for large distances from 
the wall (the bulk region) \cite{Burak04}. 
In this work, independently of the particle position with respect to the wall,
we take as the correlation hole an infinite cylinder, perpendicular to 
the wall surface, whose axis passes through this particle.
The radius $R$ of the cylinder is determined by the requirement that
the total disc surface of all cylinders $\pi R^2 N$ equals the planar interface
surface, namely
\begin{equation} \label{radius}
R^2 \, = \, \frac{q}{\pi\sigma} \, = \, 2 q^2 \ell_{\rm B} \, \mu .
\end{equation}
Note that, in units of the relevant Gouy-Chapman length $\mu$, 
$R^2/\mu^2 = 2 \,\Xi$, and up to an irrelevant prefactor,
similar choices were made in \cite{Burak04,Santangelo06}.
This means that in units of $\mu$, 
the correlation-hole radius $R$ vanishes in the PB limit,
while it goes to $\infty$ in the SC regime. Here, it can be stressed that 
$\mu$ is the relevant length scale for density gradients,
both in the WC and SC regimes, as revealed by Eqs. (\ref{PBsolution}) and (\ref{SCsolution}).

The exclusion of other particles from the cylindrical neighborhood of 
the given particle localized at $z$ modifies the electric potential 
$\psi(z)$ (\ref{elpot}) to $\psi_{\rm ch}(z) = \psi(z) - \delta\psi(z)$, where
\begin{eqnarray}
\delta\psi(z) & = & \int_0^{\infty} dz' \int_0^R d\rho 2\pi\rho
\frac{-q e n(z')}{\sqrt{(z-z')^2+\rho^2}} \nonumber \\
& = & 2\pi q e \int_0^{\infty} dz' n(z') \nonumber \\
& & \quad \times \left[ \vert z-z' \vert - \sqrt{R^2+(z-z')^2} \right] . 
\end{eqnarray}
We take $\psi_{\rm ch}(z)$ as the mean-field potential which determines
the counter-ion density via $n(z) = n_0 e^{\beta q e \psi_{\rm ch}(z)}$.
We shift the reduced correlation-hole potential 
$\phi(z) =\beta q e\psi_{\rm ch}(z)$ by a constant to fix the gauge 
$\phi(0) = 0$.
Thus, the rescaled density profile $\widetilde{n}$ is given by
\begin{equation} \label{prof}
\widetilde{n}(z) = \widetilde{n}_0 \exp\left[ \phi(z) \right] ,
\end{equation}
where the reduced potential
\begin{eqnarray} \label{phi}
\phi(z) & = & - z + \int_0^{\infty} dz' \widetilde{n}(z')  \nonumber \\ 
& & \quad \times \left( \sqrt{2\Xi+(z-z')^2} - \sqrt{2\Xi+z'^2} \right)   
\end{eqnarray}
satisfies the gauge $\phi(0)=0$ and the normalization constant 
$\widetilde{n}_0$ is determined by the electroneutrality condition 
(\ref{normalization}). 

The explicit value of $\widetilde{n}_0$ can be derived in close analogy
with the above PB treatment.
We first differentiate the $\phi$-potential with respect to $z$:
\begin{equation} \label{derphip}
\frac{d}{dz} \phi(z) =  - 1 + \int_0^{\infty} dz' \widetilde{n}(z')  
\frac{z-z'}{\sqrt{2\Xi+(z-z')^2}} .  
\end{equation}  
The integral
\begin{eqnarray}
\int_0^{\infty} dz \left( \frac{d \phi}{dz} +1 \right) \widetilde{n}(z) 
& = & \int_0^{\infty} dz \int_0^{\infty} dz' \widetilde{n}(z) \widetilde{n}(z') 
\nonumber \\ & & \quad \times  \frac{z-z'}{\sqrt{2\Xi+(z-z')^2}}
\end{eqnarray}
vanishes due to the anti-symmetric property with respect to the interchange $z\leftrightarrow z'$ of the function under integration in the rhs.
Then the equality
\begin{equation}
\int_0^{\infty} dz \frac{d \widetilde{n}}{dz} = - \int_0^{\infty} dz \widetilde{n}(z)
\end{equation}
implies the contact-value theorem
\begin{equation}
\widetilde{n}(0) \,=\, \int_0^{\infty} dz \, \widetilde{n}(z) \, =\, 1 .   
\end{equation}
We see that, as is the case within PB theory, the density normalization 
automatically ensures the validity of the contact-value theorem.
This is a nontrivial and exact property of our Coulombic system \cite{MaTT15}, that an approximate or phenomenological theory
may violate (in this respect, it is thus remarkable that PB theory
does fulfill this condition). 
None of the theories presented in \cite{Burak04} or \cite{Santangelo06}
do obey the contact theorem.
The gauge $\phi(0)=0$ fixes the normalization constant $\widetilde{n}_0=1$.
The density profile then takes the form
\begin{equation} \label{density}
\widetilde{n}(z) = \exp\left[ \phi(z) \right] ,
\end{equation}
where the reduced potential $\phi(z)$ is given by (\ref{phi}).

To summarize at this point, our key relation is (\ref{density}), supplemented by the
closure relation (\ref{phi}). The latter expresses the test-particle potential $\phi$
in terms of the mean counter-ion density, in a self-consistent fashion.

\section{Analytical results} 
\label{IV}

To begin with, it is straightforward to realize that in the 
weak-coupling limit $\Xi\to 0$, the reduced potential (\ref{phi})
takes the PB form (\ref{phiPB}).
Due to the same normalization $\widetilde{n}_0=1$, our correlation-hole 
profile (\ref{prof}) reduces to the PB one (\ref{PBfinal}).
In this section, we prove that our correlation-hole theory also provides 
the exact density profiles in the strong coupling limit, where a series
expansion is constructed to account for corrections to SC.
Then, we focus on the tail of the ionic profile, showing that it is of mean-field type,
and working out at arbitrary $\Xi$ the corresponding large-$z$ correction to the dominant tail.
All these results will be compared to numerical data in section 
\ref{VI}.

\subsection{SC limit}
In the SC limit $\Xi\to\infty$, assuming that $\widetilde{n}$ is short-ranged
(e.g., decaying exponentially) and all its moments exist, we can perform 
in Eq. (\ref{phi}) the expansion 
\begin{equation}
\sqrt{2\Xi+(z-z')^2} - \sqrt{2\Xi+z'^2} \sim 
\frac{(z-z')^2}{2\sqrt{2\Xi}} - \frac{z'^2}{2\sqrt{2\Xi}}  
\end{equation}
to obtain $\phi(z) = \phi_{\rm SC}(z) = -z$.
Inserting this one-body potential due to the surface-charge density
into (\ref{density}) reproduces the SC solution (\ref{SCsolution}).

To construct an expansion around the SC limit, we anticipate
the systematic $1/\sqrt{\Xi}$-expansion of the density profile of the form
\begin{equation}
\widetilde{n}(z) = e^{-z} \left[ 1 + \sum_{k=1}^{\infty} \frac{f_k(z)}{(2\Xi)^{k/2}} 
\right]
\end{equation}
with as-yet unknown functions $f_k(z)$.
The contact theorem (\ref{contact}) fixes the values of these functions
at the wall,
\begin{equation} \label{fcontact}
f_k(0) = 0 ,
\end{equation}
and the normalization (\ref{normalization}) fixes their integrals over $z$,
\begin{equation} \label{fnormalization}
\int_0^{\infty} dz f_k(z) = 0 .
\end{equation}
Since $\phi(z) = \ln \widetilde{n}(z)$, we have
\begin{equation}
\phi(z) = -z + \ln \left[ 1 + \sum_{k=1}^{\infty} \frac{f_k(z)}{(2\Xi)^{k/2}} 
\right] .
\end{equation}
Consequently, 
\begin{equation} \label{eq1}
\frac{d}{dz} \phi(z) = -1 +
\frac{1}{1 + \sum_{k=1}^{\infty} \frac{f_k(z)}{(2\Xi)^{k/2}}}
\sum_{l=1}^{\infty} \frac{f'_l(z)}{(2\Xi)^{l/2}} .
\end{equation}
At the same time, from (\ref{derphip}) we get
\begin{eqnarray}
\frac{d}{dz} \phi(z) & = & -1 + \int_0^{\infty} dz' e^{-z'}
\left[ 1 + \sum_{k=1}^{\infty} \frac{f_k(z')}{(2\Xi)^{k/2}} \right] \nonumber \\
& & \times \frac{(z-z')}{\sqrt{2\Xi}} \left[ 1 + 
\sum_{l=1}^{\infty} \binom{-1/2}{l} \frac{(z-z')^{2l}}{(2\Xi)^l} \right] . 
\phantom{aaa} \label{eq2}
\end{eqnarray}
Comparing the last two relations, we obtain an infinite iterative sequence 
of equations which relate $f'_l(z)$ to all $f_k(z)$ with $k\le l-1$.
It turns out that $f_k(z)$ is a polynomial of order $2k$, the absolute
term is equal trivially to zero because of the contact condition
(\ref{fcontact}).

The first correction to the SC profile reads as
\begin{equation}
f_1(z) = \frac{z^2}{2} - z .
\end{equation}
Writing formally the SC density profile plus the first correction as
\begin{equation}
\widetilde{n}(z) = e^{-z} \left[ 1 + \frac{1}{\theta} 
\left( \frac{z^2}{2} - z \right) \right] ,
\end{equation}
we have $\theta = \sqrt{2 \Xi} = 1.414 \sqrt{\Xi}$.
This has to be compared with the very accurate estimate based on the Wigner crystal
$\theta = 1.771 \sqrt{\Xi}$ \cite{Samaj11}.
A similar result $\theta\propto \sqrt{\Xi}$ was obtained in 
Ref. \cite{Hatlo10}. On the other hand, 
the finding $\theta = \Xi$ of the renormalized virial expansion 
\cite{MoreiraNetz01} fails in the dependence on $\Xi$.
Indeed, Monte Carlo simulations fully corroborate the $\theta \propto \Xi^{1/2}$ scaling
\cite{Samaj11}.

The next expansion functions read
\begin{eqnarray}
f_2(z) & = & \frac{z^4}{8} - \frac{z^3}{2} + \frac{z^2}{2} - z , \nonumber \\ 
f_3(z) & = & \frac{z^6}{48} - \frac{z^5}{8} + \frac{z^4}{8} - \frac{z^3}{6} 
- \frac{z^2}{2} - z , \nonumber \\ 
f_4(z) & = & \frac{z^8}{384} - \frac{z^7}{48} + \frac{z^5}{6} 
- \frac{17 z^4}{24} + z^3 - 3 z^2 - 3 z , \phantom{aa}
\end{eqnarray}
etc.
It is interesting that the normalization constraint (\ref{fnormalization})
is automatically ensured by respecting the contact relation (\ref{fcontact}),
which can serve as a check of algebra.
Note that, at arbitrary order of the expansion around the SC limit, the
density profile is decaying exponentially.

\subsection{Large-distance decay: asymptotic validity of PB}  
For any finite value of the coupling $\Xi$ and at asymptotically large distances from the wall 
($z\to\infty$), 
the exact density profile is
expected to exhibit the PB power-law behavior (\ref{PBsolution})
\cite{Shklovskii99,MoreiraNetz01,Chen,Santangelo06,Santos09,Varenna}, $\widetilde{n}(z)\sim 1/z^2$. 
It is worthwhile emphasizing that this power law behavior
implies that the (unscaled) counter-ion density becomes independent of
the surface charge density $\sigma$, thereby revealing a universal behavior.
An important feature of our theory is that this asymptotic behavior
indeed takes place, at variance with the approach of Ref. \cite{Burak04}.

To prove this fact, let us first assume that at large distances
\begin{equation} \label{nas}
\widetilde{n}(z) \mathop{\sim}_{z\to\infty} \frac{a}{z^2}
\end{equation}
with some positive number $a$ which might depend on $\Xi$. 
Since the positive density $\widetilde{n}$ does not exhibit divergent 
singularities, it must be bounded from above at any point $z$ by the function
\begin{equation} \label{densityineq}
\widetilde{n}(z) \le \frac{A}{(1+z)^2} ,
\end{equation} 
where $A\ge a$.
For $\Xi=0$ we can take $A=1$, while in the SC limit $\Xi\to\infty$
we have $A = 4/e = 1.47152\ldots$. The precise value of $A$ is immaterial, as long 
as it is finite.
Writing $-1$ in Eq. (\ref{derphip}) as $-\int_0^{\infty} dz' \widetilde{n}(z')$, 
the potential derivative is expressed after simple algebraic manipulations 
as follows
\begin{equation} \label{crucial}
\frac{d}{dz} \phi(z) = - 2 \int_z^{\infty} dz' \widetilde{n}(z')
- I_1(z,\Xi) + I_2(z,\Xi) ,  
\end{equation}  
where 
\begin{eqnarray}
I_1(z,\Xi) & = & \int_0^z dz' \widetilde{n}(z')  
\left[ 1 - \frac{z-z'}{\sqrt{2\Xi+(z-z')^2}} \right] , \nonumber \\  
I_2(z,\Xi) & = & \int_z^{\infty} dz' \widetilde{n}(z')  
\left[ 1 - \frac{z'-z}{\sqrt{2\Xi+(z-z')^2}} \right] .
\end{eqnarray}
Both $\widetilde{n}(z')$ and the functions in square brackets are positive.
Using the inequality (\ref{densityineq}), in the large-$z$ limit 
the integrals are bounded from above by
\begin{eqnarray}
I_1(z,\Xi) & \le & A \, \frac{\Xi+\sqrt{2 \,\Xi}}{z^2} + O\left( \frac{1}{z^3} \right) , 
\nonumber \\
I_2(z,\Xi) & \le & A \, \frac{\sqrt{2\,\Xi}}{z^2} + O\left( \frac{1}{z^3} \right) . 
\end{eqnarray}
Considering these bounds in (\ref{crucial}), it holds that
\begin{equation} \label{derequation}
\frac{d}{dz} \phi(z) = - 2 \int_z^{\infty} dz' \widetilde{n}(z') +
O\left( \frac{1}{z^2} \right) .
\end{equation}  
Since $\phi(z) = \ln \widetilde{n}(z)$, the asymptotic formula (\ref{nas}) 
implies that $\phi'(z) \mathop{\sim}_{z\to\infty} -2/z$.
Inserting this asymptotic relation together with (\ref{nas}) into
Eq. (\ref{derequation}), one gets $a=1$.
Consequently, at any finite value of the coupling $\Xi$, the asymptotic 
large-distance behavior of the density profile is exactly of PB type,
as was expected.
This property is confirmed also by a numerical treatment of our correlation-hole
equations in the next section.

\subsection{Subleading asymptotic correction}

It is possible to go one step further and to compute the large-$z$
correction to the mean-field asymptotics (large-$z$ analysis at fixed $\Xi$).
We use the electroneutrality condition (\ref{normalization}) to rewrite
the correlation-hole relation (\ref{derphip}) as follows
\begin{equation}
\frac{d}{dz} \phi(z) = - 2 \int_z^{\infty} dz' \widetilde{n}(z') + I(z,\Xi) , 
\label{eq:36}
\end{equation}
where
\begin{equation}
I(z,\Xi) = \int_0^{\infty} dz' \widetilde{n}(z') \left[ 
\frac{z-z'}{\sqrt{2\Xi+(z-z')^2}} - \frac{z-z'}{\vert z-z'\vert} \right] .  
\label{eq:37}
\end{equation}
To proceed, we change variables $z\to u=(1+z)^{-1}$, and perform 
a small $u$ expansion in Eqs. (\ref{eq:36}) and (\ref{eq:37}).
Using the fact that $I(u,\Xi) = -\Xi u^2 + o(u^2)$, writing 
$n(u) = u^2 + \Delta n(u,\Xi)$ and keeping in mind that $\Delta n$
is $o(u^2)$ but not necessarily $O(u^3)$, we get
\begin{equation}
-\frac{\partial^2 \Delta n}{\partial u^2} + \frac{2}{u} \frac{\partial \Delta n}{\partial u}
+ 2\Xi u + o(u) = 0,
\end{equation}
from which the 
correction to the PB asymptotics follows:
\begin{equation}
\widetilde{n}(z) \,\sim \, \frac{1}{(1+z)^2} - \frac{2}{3} \,\Xi\, \frac{\log(1+z)}{(1+z)^3}.
\label{eq:corr}
\end{equation}
As the exact loop-derived correction,  it is of order 
$\Xi$ and decays at large $z$ like $z^{-3}\log(z)$ \cite{Attard88,Podgornik90,Netz00}.
Yet, our $-2/3$ prefactor for the correction in Eq. (\ref{eq:corr}) is not equal to that reported in \cite{Netz00}, which is
$-1$. We mention here that repeating the analysis of \cite{Netz00} lead us to a corrected prefactor
$-1/2$, closer to the present $-2/3$.

We shall see below that the predicted correction is indeed found in the numerical 
treatment of our self-consistent scheme. For large $\Xi$ however, it becomes
practically impossible to reach the relevant distance range, and another
contribution preempts that in Eq. (\ref{eq:corr}), for the range of available
distances. This is discussed further below.

\section{Numerical results} \label{VI}

\subsection{The methods}

The correlation-hole integral equation for the rescaled density profile 
$\widetilde{n}$, given by  Eqs. (\ref{phi}) and (\ref{density}),
bears some similarities with the nonlinear PB formulation.
Solving it numerically is straightforward. In practice, an efficient numerical 
scheme was found to be the following. 
Rescaled distances $z$ are first mapped onto a variable $x=(1+z)^{-1/2}$,
such that $x\in [0,1]$. The resulting equations for $\phi(x)$ is then discretized
on a regular grid with $N$ points ($N$ up to $2\times 10^5$). We initialize the density
to be of PB form, meaning that $\widetilde n(x) = x^4$ (which results in an 
improved convergence), before an iterative resolution.
Convergence is typically achieved in 100 iterations if fine properties are sought.
It is important here to emphasize that from a computational point of view,
the resolution of our self-consistent equation is significantly faster and
more convenient than the test charge approach \cite{Burak04}, or the
theory of Santangelo \cite{Santangelo06}.

In parallel, we have performed a number of Monte Carlo simulations
in a quasi-2D geometry. Ewald summation techniques corrected for quasi-2D-dimensionality
allow to account for long-range electrostatic interactions 
(see e.g. \cite{Berkowitz,Mazars,JPCM18}). The Monte Carlo 
results provide the correct reference behavior of our system of point ions
in the vicinity of a charged plate.

\subsection{Comparison to Monte Carlo results}

\begin{figure}[tbp]
\begin{center}
\includegraphics[width=0.45\textwidth,clip]{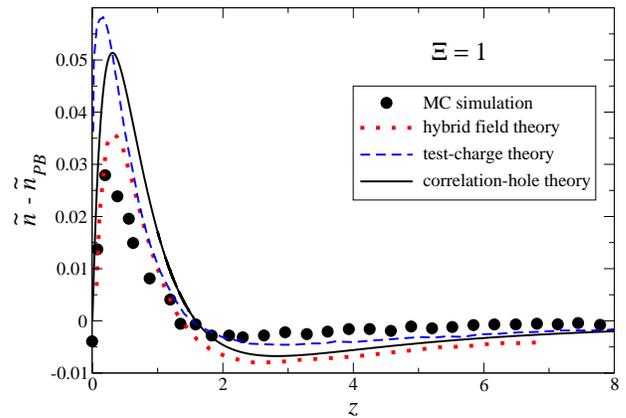}
\caption{Deviation from the PB density profile, 
$\widetilde{n}-\widetilde{n}_{\rm PB}$, as a function of the dimensionless 
distance $z$ for the coupling constant $\Xi=1$.
Symbols correspond to the results of Monte Carlo simulation,
the dashed curve is for the test-charge theory of Ref. \cite{Burak04}, the dotted curve 
is for the approach of Ref. \cite{Santangelo06}, and solid curve
shows the present correlation-hole approach.}
\label{fig:dens1}
\end{center}
\end{figure}

\begin{figure}[tbp]
\begin{center}
\includegraphics[width=0.45\textwidth,clip]{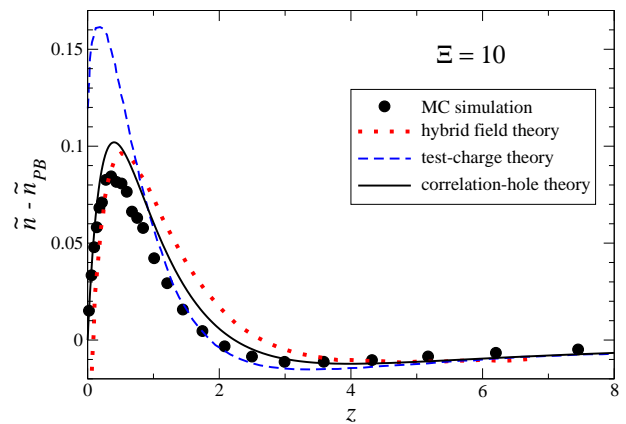}
\caption{Same as Fig. \ref{fig:dens1} for the coupling $\Xi=10$. }
\label{fig:dens10}
\end{center}
\end{figure}

\begin{figure}[tbp]
\begin{center}
\includegraphics[width=0.45\textwidth,clip]{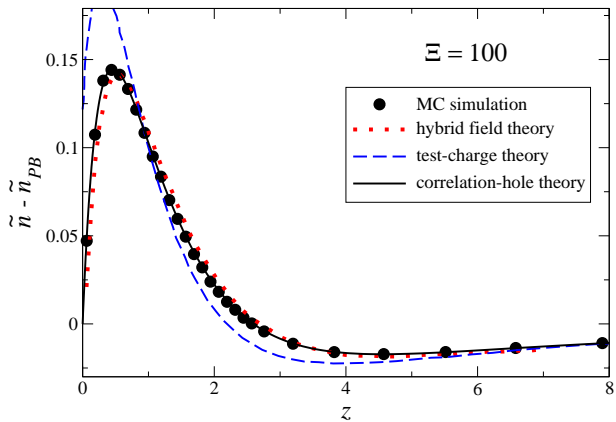}
\caption{Same as Fig. \ref{fig:dens1} for $\Xi=100$.}
\label{fig:dens100}
\end{center}
\end{figure}

The numerical results for the deviations from PB profiles, 
$\widetilde{n}-\widetilde{n}_{\rm PB}$
are presented for the coupling constants $\Xi=1$, $\Xi=10$ and $\Xi=100$
in Figs. \ref{fig:dens1}, \ref{fig:dens10} and \ref{fig:dens100} respectively.
Our MC simulations are compared to the test-charge theory \cite{Burak04}, to the hybrid field theory
of Ref. \cite{Santangelo06} where long and short distances are treated separately  and 
to the present correlation-hole approach.
We see that for the small value $\Xi=1$, the accuracy of the test-charge and 
correlation-hole theories is comparable. The hybrid field theory of
Ref. \cite{Santangelo06} (which is solved at the expense of 
enhanced technical complexity) fares better at short distances, but worse for $z>2$.
For intermediate $\Xi=10$, the accuracy of our
approach is better.
For relatively large $\Xi=100$, our solid curve practically passes through 
MC data.
The accuracy of our results improves upon increasing $\Xi$.  

\begin{figure}[htbp]
\begin{center}
\includegraphics[width=0.45\textwidth,clip]{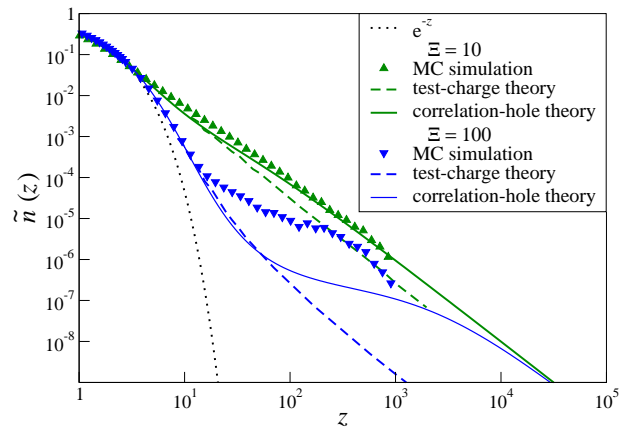}
\caption{Large-distance counter-ion densities for $\Xi=10$ (in green) and $\Xi=100$ (in blue). Monte Carlo data (symbols)
are compared to the correlation-hole results (solid curves) and those of the test-charge theory 
of Ref. \cite{Burak04} (dashed lines). The dotted line is for the SC limiting behavior $\Xi\to \infty$.}
\label{fig:I3}
\end{center}
\end{figure}

For the tail of the ionic profile, at larger distances than those in the
previous graphs, we see in Fig. \ref{fig:I3} that the correlation-hole 
picture captures qualitatively the departure from SC behavior, although in a distance range
that is not close enough to the
charged plate. Yet, the test charge theory fails in getting the qualitative trend.
For $\Xi=100$, the MC result clearly follow the exponential profile at $\widetilde z < 10$ \cite{MoreiraNetz01},
then crosses over to a longer range decay, following a trend that is reminiscent of that 
observed within the correlation-hole approach (same shape in the log-log plot 
presented). Observing properly the PB algebraic tail in $1/z^2$, with MC at $\Xi>100$, would require
significantly larger systems, a relative accuracy on the profiles better than $10^{-6}$,
and is beyond our scope. For this reason and in order to 
study nevertheless the crossover to mean-field, we will in the remainder relinquish MC method and
focus on the self-consistent treatment, which is considerable simpler to solve.

\subsection{Discussion of asymptotic features}

\begin{figure}[htbp]
\begin{center}
\includegraphics[width=0.45\textwidth,clip]{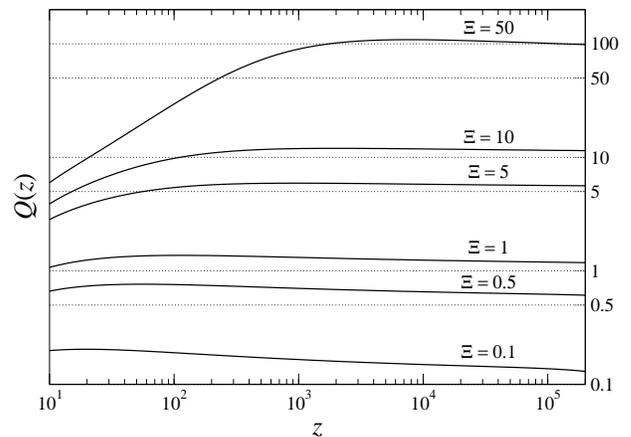}
\caption{Plot of $Q(z)$ as defined in Eq. (\ref{eq:corrQ}), vs distance to the charged wall.
Eq. (\ref{eq:corr}) predicts that $Q$ asymptotically tends to $\Xi$, indicated the by horizontal dotted 
lines.}
\label{fig:I2}
\end{center}
\end{figure}

We wish to investigate the behavior of ionic density at large distances,
to first test the relevance of the correction worked out in Eq. (\ref{eq:corr}),
but also to discuss the crossover to the mean-field regime. 
Fig. \ref{fig:I2} extracts the correction to the PB profile,
and compares it to the predicted functional form in $\Xi \,\log(1+z)/(1+z)^3$.
This is achieved through the computation of the following quantity:
\begin{equation}
Q(z) = \frac{ n(z) - (1+z)^{-2} }{   -2/3 \, \log(1+z) / (1+z)^3 }.
\label{eq:corrQ}
\end{equation}
It is observed that for $\Xi<10$, $Q$ saturates at large distance close to the expected
value $\Xi$. For $\Xi=50$ (and higher), the range of distances probed does not allow to
reach large enough $z$ to observe the phenomenon. 

\begin{figure}[htbp]
\begin{center}
\includegraphics[width=0.45\textwidth,clip]{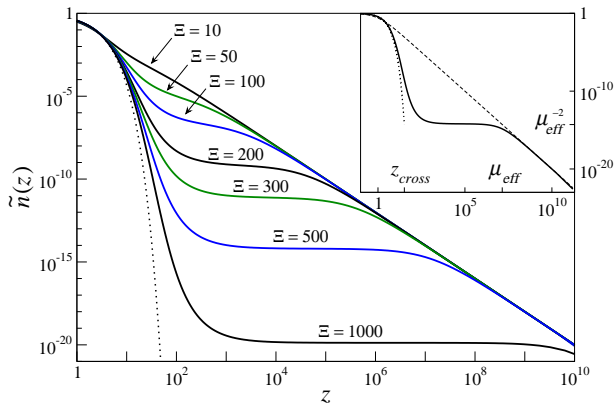}
\caption{Scenario for the density large-distance  asymptotics.
The SC limiting behavior on the left hand side is displayed with the dashed line.
The inset shows the crossover distance $z_{\text{cross}}$ 
and the effective Gouy-Chapman length $\mu_{\text{eff}}$ for $\Xi = 500$. 
}
\label{fig:I4}
\end{center}
\end{figure}	

For $\Xi>50$, the large-$z$ density profile exhibits a new property, that is only 
beginning to emerge in Fig. \ref{fig:I3}. This is illustrated in Fig. \ref{fig:I4}:
the  expected exponential SC regime at short $z$ and mean-field tail at large $z$ 
are connected by a plateau, starting at the crossover length $\zcross$, 
where the density is quasi-constant. To be more specific,
the existence of a plateau followed by a $z^{-2}$ decay is precisely the PB prediction,
with an effective Gouy-Chapman length $\mueff$, and a density
\begin{equation}
\widetilde n(z) \, = \, \frac{1}{(z+\mueff)^2} .
\label{eq:mueff}
\end{equation}
Thus, for $z<\mueff$ (but $z>\zcross$), the density profile is flat, while 
for $z>\mueff$, it decays algebraically. Keeping in mind that by its definition in Eq. (\ref{eq:gc}), 
a Gouy-Chapman length scales like the inverse plate charge, it is natural
to expect $\mueff$ to largely exceed the bare Gouy-Chapman length. Indeed,
the PB-like profile sets in for $z>\zcross$, and subsumes all nonlinear screening effects
at work for $0<z<\zcross$
into an effective plate surface charge, thus significantly smaller than $\sigma$.

\begin{figure}[htbp]
\begin{center}
\includegraphics[width=0.45\textwidth,clip]{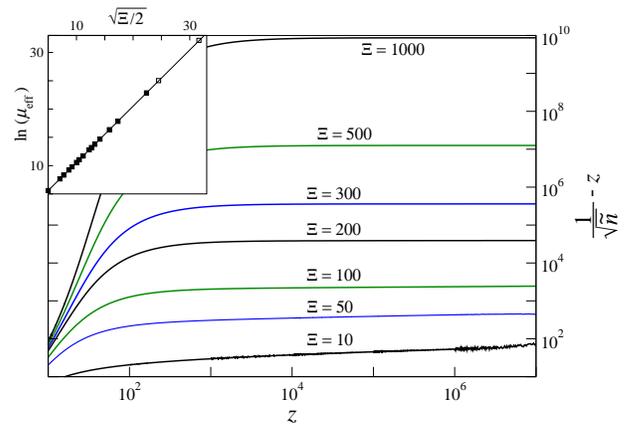}
\caption{Extraction of the effective Gouy-Chapman length $\mu_{\text{eff}}$, from the plot of 
$1/\sqrt{\widetilde n} - z$, for $\Xi$ between 10 and 1000. The plateau reached defines $\mu_{\text{eff}}$.
The inset shows how the resulting effective length depends on the coupling parameter.
The line has slope 1.}
\label{fig:I5}
\end{center}
\end{figure}

It can be noted that the large-$z$ expansion of Eq. (\ref{eq:mueff}) yields $\widetilde n \sim 1/z^2 -2 \mueff/z^3$.
The resulting correction to the $1/z^2$ tail is of smaller order than 
the term in $\log z / z^3$ stemming from Eq. (\ref{eq:corr}). Hence, the value of $\mueff$ 
cannot be simply extracted from the asymptotic tail of the profile,
but at smaller distances, where Eq. (\ref{eq:mueff}) is relevant \cite{rque10}. The plateau seen in Fig. \ref{fig:I4}
illustrates this point: for $z>\zcross $, Eq. (\ref{eq:mueff}) states that 
$\widetilde n^{-1/2}$ increases linearly with distance, so that the quantity displayed in Fig.
\ref{fig:I5} offers a convenient measure of the effective Gouy-Chapman length.
It can be observed in Fig. \ref{fig:I5} that for $\Xi=10$, one cannot properly 
extract a $\mueff$, which is consistent with the data in Fig. \ref{fig:I3}
(absence of a well defined plateau).
The inset of Fig. \ref{fig:I5}, where the line shown has equation $y=x+0.62$,
then indicates that $\mueff$ changes with $\Xi$ as 
\begin{equation}
\log \mueff \sim \sqrt{ \frac{\Xi}{2} } \, + \, \hbox{cst} .
\end{equation}
This in turn sets the crossover distance to be
\begin{equation}
\zcross \, \sim \, \sqrt{2\,\Xi},
\end{equation}
by equating $e^{-z}$ with $1/\mueff^2$ at $\zcross$. It does not come as a surprise to 
recover here the value of the correlation-hole size \cite{Burak04,Santangelo06}, see Eq. (\ref{radius}) which reads
$\widetilde R^2 = 2\, \Xi$.
The effective length $\mueff$ diverges with $\Xi$, such that $\log\mueff$ is linear
in $\sqrt{\Xi}$,
a conclusion also reached
in \cite{Santangelo06}. 
Large values of $\mueff$ were observed numerically as well 
in the case of counter-ions around charged cylinders \cite{Mallarino13}.

\begin{figure}[htbp]
\begin{center}
\includegraphics[width=0.45\textwidth,clip]{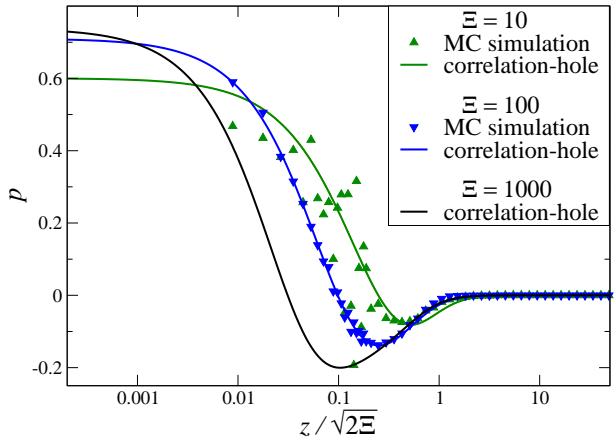}
\caption{Implementing our  mean-field probe. The vanishing of the local pressure $p(z)$,
as defined in Eq. (\ref{eq:probe}), signals the PB regime.  On the $x$-axis, distances
have been rescaled by $\zcross = \sqrt{2 \Xi}$. A system with $\Xi <1$ exhibits 
a flat $p=0$ curve, since mean-field holds at all distances. For large enough $\Xi$,
$p$ starts at 3/4 for small $z$ (since $\widetilde n=\exp(-z)$  locally holds), then reaches a minimum
value close to $-1/4$, before vanishing on a scale $\zcross$. Symbols are for MC, and the curves
for the correlation-hole theory. 
}
\label{fig:Ilast}
\end{center}
\end{figure}

Finally, we present an operational way to decide when a system with an arbitrary $\Xi$ is in the mean-field regime.
The idea is to take advantage of 
the fact that the stress tensor is divergence-free \cite{ET2001}.
For mean-field theories, this yields an extended contact theorem (not only at $z=0$, but at any $z$). 
In the present planar geometry,
this means that, using dimensionless quantities
\begin{equation}
p(z) \, \equiv \, \widetilde n(z) \,-\,\frac{1}{4}\, [\phi'(z)]^2 \, = \, 0 .
\end{equation}
To check for that identity with numerically obtained results, one could 
compute the correct potential $\phi$, from integrating the charge density.
However, keeping in mind that we seek here a mean-field probe, it is more convenient
to assume $\phi = \log\widetilde n$ and we arrive at
\begin{equation}
p(z) \, \equiv \, \widetilde n(z) \,-\,\frac{1}{4}\, [\partial_z \log \widetilde n]^2 \, = \, 0 .
\label{eq:probe}
\end{equation}
Deviations of $p(z)$  from 0 provide a (sufficient) condition for mean-field violation. 
The fact that $p=0$ within a mean-field treatment is a consequence of the contact
theorem, that reads $p(0)=0$. It indicates that the pressure vanishes in our setting
(single plate problem, corresponding to a two-plate in interaction, in the limit where 
inter-plate distance is infinite).
Figure \ref{fig:Ilast} corroborates the existence of a PB tail, at large enough distances.
Yet, a word of caution is in order here. It can rightfully be argued that a quantity
such as $p(z)$ may only distinguish exponential profiles from algebraic ones, but that any 
density of the type $\widetilde n \propto (\mueff + z)^\alpha$ yields $p\to 0$ for all $\alpha>0$,
and not only $\alpha=2$. A possible solution would be to consider the ratio of the two terms
subtracted
in \eqref{eq:probe}, rather than their difference; the ratio goes to a constant for the PB
behavior only ($\alpha=2$). However,
this has a drawback: it amplifies the contribution of any residual exponential tail in the density, and requires 
larger distances to qualify the density as PB-like. A point to keep in mind though 
is that our probe \eqref{eq:probe} is more interesting for a two plate system
where the real (e.g. Monte Carlo) pressure $P$ is non-vanishing, rather than for the one plate 
situation. Indeed, in such a case, comparing $p(z)$ to $P$ can be viewed as signaling 
the mean-field regime. 


\section{Concluding remarks} \label{VII}

We have studied a system of identical counter-ions near a wall carrying
a uniform surface charge density, in thermal equilibrium.
This is probably the simplest model of the electrical double layer,
depending only on one parameter, the coupling constant $\Xi$.
It provides an interesting test-bench, since both Weak Coupling (WC) and Strong Coupling 
(SC) limits are known.

We have proposed a method which combines physical ideas from both WC 
and SC regimes.
From the WC side, the particle density is determined by the Boltzmann factor 
of the mean potential.
From the SC side, there is a cylindrical correlation hole around each 
particle, inaccessible to other particles, which modifies the value of 
the mean potential.
The theory is simple by its construction and leads to a nonlinear integral 
equation, similar to the PB one, which converges quickly in an iterative scheme. 

Remarkably, all exact constraints are respected 
by our correlation-hole theory, for all coupling constants.
The contact theorem for the particle density at the wall holds.
The WC and SC limits are reproduced as well, and the correction to the SC limit $\Xi\to\infty$ is
proportional to $1/\sqrt{\Xi}$, in accordance with recent approaches
and MC simulations.
For large distances from the wall and at arbitrary $\Xi$, the algebraic mean-field density profile is recovered.
Moreover, we showed that the corresponding subleading correction, in $\Xi \log z / z^3$,
is of the same form as found in a loop-wise field theoretic treatment of fluctuations 
beyond Poisson-Boltzmann \cite{Netz00}.
Focusing on the approach to mean-field behavior at large distances, we 
showed that beyond a crossover distance $\zcross$ (coinciding with the hole size), 
the density takes a Poisson-Boltzmann form. 
This allows to define 
an effective Gouy-Chapman length to describe the density tail. In units of the 
bare length $\mu$, it behaves as $\mueff \propto \exp(\sqrt{\Xi/2})$,
and quickly grows with $\Xi$. This is a signature of efficient nonlinear screening,
leading to a small effective surface charge for the plate, as far as its large
scale potential is concerned. Introducing a ``mean-field probe'', $p(z)$ in Eq.
(\ref{eq:probe}), we recover the results of a direct analysis of the numerical
profiles.

\begin{figure}[htbp]
\begin{center}
\includegraphics[width=0.45\textwidth,clip]{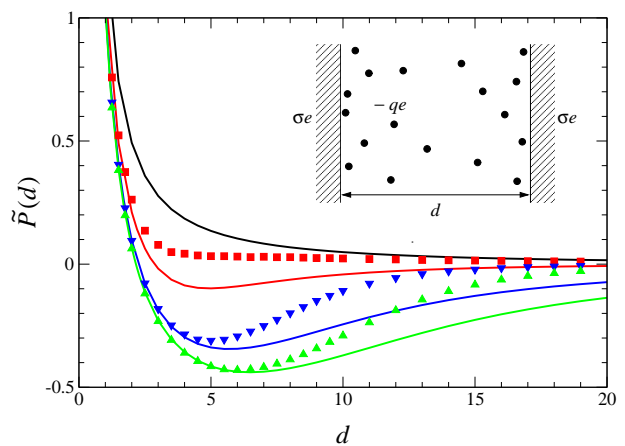}
\caption{Interplate pressure versus rescaled distance, for $\Xi = 1$, 10, 50 and 100 (from top to bottom).
Monte Carlo results (symbols) are compared to the prediction of the correlation-hole theory (lines). 
The rescaled pressure is defined as $\widetilde P = P / (k_B T 2\pi \ell_B \sigma^2)$,
and is measured from the contact theorem.}
\label{fig:I6}
\end{center}
\end{figure}

For the sake of completeness, we also considered the situation of two
parallel uniformly charged plates (surface charge density $\sigma e$), at distance $d$,
sandwiching a slab of counter-ions. 
There, an ambiguity arises when enforcing the idea of a correlation hole. 
Indeed, we have to distinguish between the two limits $d\to\infty$ and $d\to 0$.
Accepting the cylinder form of the correlation hole, the cylinder radius is 
given by formula (\ref{radius}) if $d\to\infty$, i.e. 
$\sigma \pi R_{\infty}^2 = q$, and by $2 \sigma \pi R_0^2 = q$ if $d\to 0$.
A possible, $d$-dependent interpolation formula for the correlation-hole size
might be relevant, but for simplicity, we took the same prescription as 
in the one-plate case, Eq. (\ref{radius}). The alternative choice turned out to be slightly
worse. 
The equation of state 
of this system, as measured in Monte Carlo simulations, is reported 
in Fig. \ref{fig:I6}. To test our correlation-hole approach (accurate at both 
small and large couplings), we concentrate in Figure \ref{fig:I6} on
intermediate coupling strengths, where the phenomenon of like-charge
attraction sets in \cite{Messina09,Levin02,MoreiraNetz00,Varenna}. 
We see that the qualitative features of the pressure are well captured,
with an agreement that is quantitative for small distances, up to the range 
where like-charge attraction is maximal (minimum of the pressure). The asymptotic 
decay to vanishing pressure then takes place over too large distances, as compared to
MC. The correlation-hole idea there overestimates the SC non-mean-field features;
correcting for this deficiency is left for future work. Yet, it is noteworthy that 
the present theory captures here also a number of exact features. Not only is the proper
equation of state recovered when $\Xi\to 0$ and $\Xi \to \infty$, but also,
the pressure minimum arises at $z\propto \Xi^{1/4}$, as found in Monte Carlo
simulations \cite{Chen,Samaj11}.

\begin{acknowledgments}
It is a pleasure to dedicate this work to Daan Frenkel as a tribute to his exceptional 
human and scientific qualities (depth, vista, swiftness), be it as a mentor or a colleague. 
To counter-quote Churchill, Daan is a modest man, who does not have much to be modest about.
The support received from VEGA Grant No. 2/0003/18 is acknowledged. 
The work was funded by the European Union's Horizon 2020 research and innovation programme under 
ETN grant 674979-NANOTRANS. M.T. acknowledges financial support by the Swedish Research Council (621-2014-4387).
\end{acknowledgments}


\begin{thebibliography}{10}
\bibitem{rque10}
Yet, as alluded to above, reaching the $z$-range where the log correction is
relevant becomes in practice impossible when increasing $\Xi$. For instance, we would need 
$z>10^6$ for $\Xi=50$ and already $z>10^{29}$ for $\Xi=100$. Thus, the log correction can only be probed 
at $\Xi<50$.
\end{thebibliography}
\end{document}